\documentclass[preprint,showpacs,preprintnumbers,amsmath,amssymb]{revtex4-1}
\usepackage{graphicx}
\usepackage{dcolumn} 
\usepackage{bm} 
\usepackage{longtable}

\begin{document}

\title{Response of the topological surface state to surface disorder in TlBiSe$_2$}

\author{Florian Pielmeier$^{1}$, Gabriel Landolt$^{2,3}$, Bartosz Slomski$^{2,3}$, Stefan Muff$^{3,8}$ Julian Berwanger$^1$, Andreas Eich$^{4}$, Alexander A. Khajetoorians$^{4,9}$, Jens Wiebe$^{4}$, Ziya S. Aliev$^{5,6}$, Mahammad B. Babanly$^{5,6}$, Roland Wiesendanger$^{4}$, J\"urg Osterwalder$^{2}$, Evgeniy V. Chulkov$^{7}$, Franz J. Giessibl$^{1}$, J. Hugo Dil$^{2,3,8}$}
\address{
$^{1}$Institute of Experimental and Applied Physics, Universit\"at Regensburg, D-93040 Regensburg, Germany
\\
$^{2}$Physik-Institut, Universit\"at Z\"urich, Winterthurerstrasse 190, 
CH-8057 Z\"urich, Switzerland 
\\ 
$^{3}$Swiss Light Source, Paul Scherrer Institut, CH-5232 Villigen, 
Switzerland 
\\
$^{4}$Department of Physics, Universit\"at Hamburg, D-20355 Hamburg, Germany
\\
$^{5}$General and Inorganic Chemistry Department, Baku State University, AZ1148 Baku,  Azerbaijan
\\
$^{6}$Institute of Chemical Problems, Azerbaijian National Academy of Science, AZ1143 Baku, Azerbaijian
\\
$^{7}$Donostia International Physics Center (DIPC) and CFM-MPC, Centro Mixto CSIC-UPV/EHU,
Departamento de Fisica de Materiales, UPV/EHU, 20080 San Sebastian, Spain
\\
$^{8}$Institut de Physique de la Mati\`ere Condens\'ee, Ecole Polytechnique
F\'ed\'erale de Lausanne, CH-1015 Lausanne, Switzerland
\\
$^{9}$Institute for Molecules and Materials, Radboud University Nijmegen, 6500 GL Nijmegen, The Netherlands}

\date{\today}

\begin{abstract}
Through a combination of experimental techniques we show that the topmost layer of the topological insulator TlBiSe$_2$ as prepared by cleavage is formed by irregularly shaped Tl islands at cryogenic temperatures and by mobile Tl atoms at room temperature. No trivial surface states are observed in photoemission at low temperatures, which suggests that these islands can not be regarded as a clear surface termination. The topological surface state is, however, clearly resolved in photoemission experiments. This is interpreted as a direct evidence of its topological self-protection and shows the robust nature of the Dirac cone like surface state. Our results can also help explain the apparent mass acquisition in S-doped TlBiSe$_2$.
\end{abstract}


\maketitle
\section{Introduction}
Topological insulators (TIs) constitute a novel class of materials that has received a large amount of attention over recent years \cite{Hasan:2010,Qi:2011}. The main reason for this strong scientific interest is the presence of metallic surface states with a helical spin structure on the surface of a semiconducting bulk material, which renders them a possible candidate for spintronics applications \cite{Wolf:2001}. However, spin-polarized surface states are not a unique characteristic of topological insulators. Spin-split states have been found on the surface of a variety of systems which do not belong to this class of materials, i.e., topologically trivial Rashba systems \cite{LaShell:1996,Hofmann:2006,Dil:2009R}. The truly unique property of the surface states of topological insulators is their so-called topological protection; they can not be destroyed by perturbations that do not break time reversal symmetry. Within a simplified model it is often suggested that this protection is caused by the spin structure which suppresses backscattering events as this would require a, highly improbable, spin flip \cite{Qi:2011}. Although this simplification is certainly valid for the one-dimensional edge states of two-dimensional (2D) TIs \cite{Konig:2007}, the additional phase-space available for scattering for the 2D surface states of 3D TIs calls for a protection mechanism different from avoided backscattering. Indeed, scanning tunneling spectroscopy experiments on both topological insulators and topologically trivial materials with spin-polarized states reveal similar scattering rules around defects \cite{Roushan:2009,Alpichshev:2010,Honolka:2012}. In recent photoemission experiments on the topological insulator Bi$_2$Se$_3$ it was found that even after mild ion sputtering the topological surface state was no longer visible \cite{Hatch:2011}. Later theoretical considerations verified this behaviour and suggested that the state actually moved to the next quintuple layer \cite{Schubert:2012}. This indicates that the real protection mechanism of the topological surface state is just like its unique spin structure a consequence of the transition between phases of different topology at the edge of a TI \cite{Hasan:2010}.

\begin{figure}
	\centering
		\includegraphics[width=0.98\textwidth]{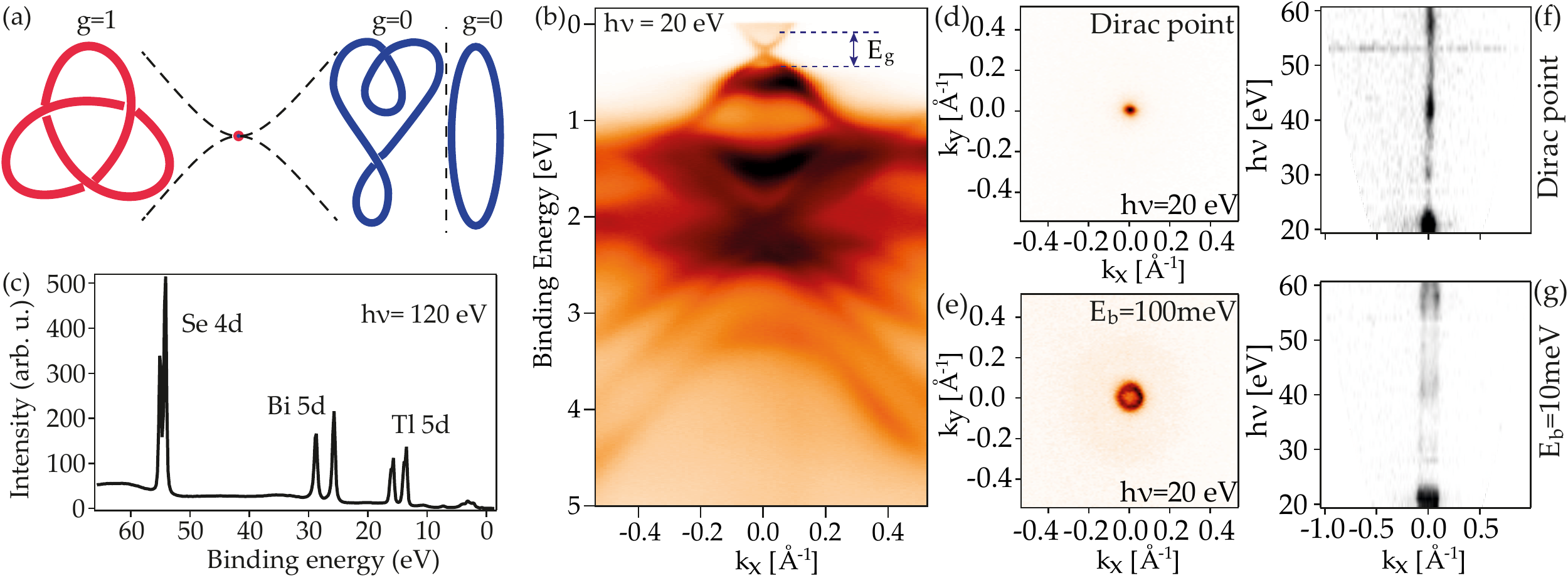}
	\caption{(a) Illustration of a topological interface by knots. The trifoil knot on the left can not be transformed into the unknot on the right without going through a singularity. (b) ARPES band map of TlBiSe$_2$ along the $\Gamma$-M direction measured at a photon energy of $20\,\mathrm{eV}$. The dashed blue lines indicate the bulk band gap $E_g$. (c) Core level XPS of TlBiSe$_2$ obtained at normal emission and a photon energy of $120\,\mathrm{eV}$. (d) and (e) Constant energy surfaces obtained at a photon energy of $20\,\mathrm{eV}$ and a binding energy of $280\,\mathrm{meV}$ corresponding to the Dirac point (d) and $100\,\mathrm{meV}$ (e). (f) and (g) Photon energy dependent scans at constant $k_y$ and a binding energy corresponding to the Dirac point (f) and $10\,\mathrm{meV}$ (g).}
	\label{fig1}
\end{figure}

Intrinsic to the definition of topological classes is that it is impossible to go from one class to another through continuous deformations. As illustrated by the knots with different topology in Figure  \ref{fig1} (a), to go from one topological class to another the system has to go through a singularity. In the electronic structure of a material the topological class, or genus $g$, is defined by the number of parity inversions in the bulk band structure \cite{Zhang:2009}. A non-trivial ($g=1$) band structure has an odd number of parity inversions, whereas a trivial band structure ($g=0$) has an even number of parity inversions. At the interface between the two systems the band gap must thus close and re-open again, which leads to the formation of an interface state \cite{Shockley:1939}. Therefore at the transition between two regions of different topological classes an interface state must exist \cite{Hasan:2010}. Here we show by a combination of angle-resolved photoemission (ARPES), scanning tunneling microscopy (STM), and atomic force microscopy (AFM) that the protection mechanism of the topological surface states is most likely based on moving away from regions with high defect density due to the fact that these regions obtain a trivial topology. 

\section{Experimental Methods}
Single crystals of TlBiSe$_2$ were grown from high purity elements using a Bridgman method. The bulk crystalline quality was checked using X-ray diffraction. All the samples used at the different facilities and different techniques originate from the same batch of crystals. Oriented single crystals were glued on the respective sample holders and cleaved in ultrahigh-vacuum (UHV) by knocking off a top-post. All presented results are reproduced for a large number of cleaves and show no clear dependence on cleaving temperature.\\
The photoemission experiments were performed using the spin-polarized ARPES end station COPHEE \cite{Hoesch:2002} as well as the high resolution ARPES end station at the Swiss Light Source  using linearly (p) polarised light at a sample temperature of $20\,\mathrm{K}$ and a base pressure better than $2 \cdot10^{-10}\,\mathrm{mbar}$. Samples were cleaved at $20\,\mathrm{K}$, $60\,\mathrm{K}$ and room temperature (RT).\\
STM experiments were performed in an Omicron multichamber UHV system with a base pressure below $1 \cdot10^{-10}\,\mathrm{mbar}$ using a home-built variable temperature STM. Both the tip, electrochemically etched from polycrystalline W wire, and sample were cooled by a Cryovac continuous flow He cryostat to $T = 30\,\mathrm{K}$. STM topography images were taken in constant current mode at a tunneling current $I$ with the bias voltage $U$ applied to the sample. The samples were cleaved at $\approx 150\,\mathrm{K}$ and RT.\\
The AFM experiments were performed with an Omicron low temperature combined STM/ AFM system operated in UHV at a temperature of $4.4\,\mathrm{K}$. The microscope is equipped with a qPlus sensor \cite{Giessibl:2000}; again W is used as tip material. The bias voltage is applied to the sample. For AFM operation, the frequency modulation mode is utilized \cite{Albrecht:1991}. Here, the oscillation amplitude $A$ (typically $A=50\,\mathrm{pm}$) is kept constant and the frequency shift $\Delta f$ of the cantilever, which is a measure of the force gradient between tip and sample, is monitored. Samples were cleaved at RT.\\
The RT STM measurements were carried out on a home-built STM/AFM system in UHV at a pressure of $2 \cdot10^{-10}\,\mathrm{mbar}$. QPlus sensors with W tips are used and the bias voltage is applied to the tip. Samples were again cleaved at RT.

\section{Results and Discussion}
Figure \ref{fig1} (b) shows ARPES data of the well established Dirac cone and a larger range of the valence band of the (001) surface of TlBiSe$_2$  \cite{Kuroda:2010,Xu:2011,Sato:2011}. Although this Dirac cone and the associated spin structure clearly establish TlBiSe$_2$ as a topological insulator we would like to draw the focus not only to the presence of this state, but also to the absence of any other surface states within the bulk band gap, and also at higher binding energies. Although density functional theory (DFT) calculations predict the presence of additional trivial surface states regardless of the exact surface termination  \cite{Eremeev:2011Tl}, no ARPES experiment, including ours, has been able to reproduce these states. This is in stark contrast, e.g., to Bi$_2$Se$_3$ where the topological surface state is found to coexist with other surface states \cite{Bianchi:2010,King:2011,Valla:2012,Benia:2011,Scholz:2012}. In order to exclude photoemission matrix element effects as the reason for the missing observation of trivial surface states, we also scanned along the perpendicular momentum direction as shown in the constant energy maps in Figure \ref{fig1} (d) and (e). Furthermore we performed photon energy dependent measurements (Figure \ref{fig1} (f) and (g)) and did not observe any additional surface states in the bulk band gap throughout the full energy range. As will be discussed below the absence of these trivial surface states and the well defined line shape of the topological state are indicative of the topological self-protection. The x-ray photoelectron spectroscopy (XPS) data in Figure \ref{fig1} (c) shows the chemical purity of the sample and gives a first hint of the surface structure through the observed double peak of the Tl 5d core levels, which reveals two types of environments for the Tl atoms implying that the surface is formed by Tl atoms \cite{Kuroda:2013}.

\begin{figure}
	\centering
		\includegraphics[width=0.98\textwidth]{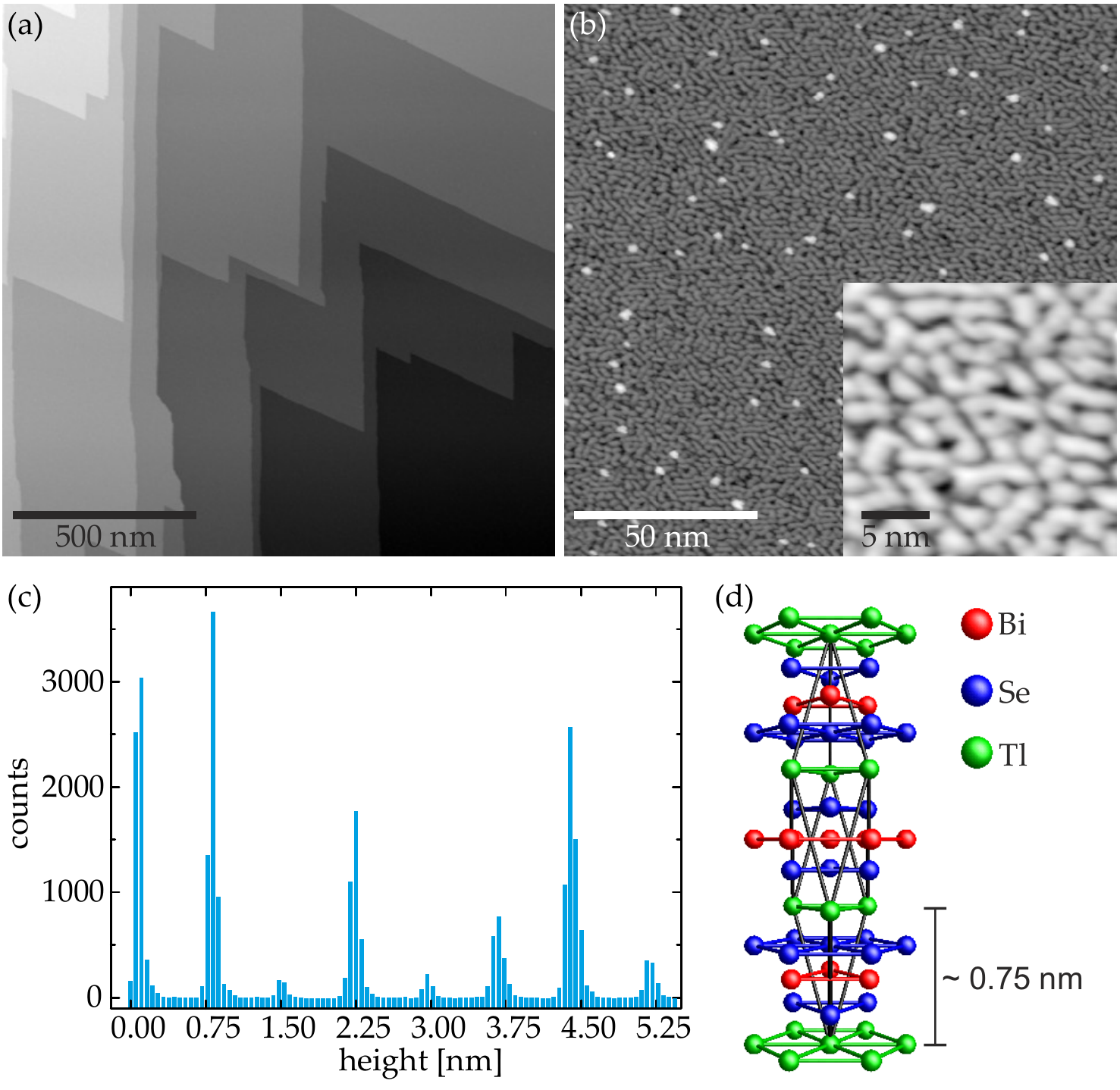}
	\caption{(a) $1.5\times1.5\,\mathrm{\mu m^2}$ constant current STM image of cleaved TlBiSe$_2$. (b) Zoomed images, resolving a disordered structure ($150\times 150\,\mathrm{nm^2}$, inset: $20\times 20\,\mathrm{nm^2}$). (c) Histogram of step height distribution in (a). The peaks are all spaced by multiples of the Tl-Tl distance of about $0.75\,\mathrm{nm}$. (d) Crystal structure model of TlBiSe$_2$ \cite{Eremeev:2011Tl}. Imaging parameters: (a,b) $I=100\,\mathrm{pA}$, $U=500\,\mathrm{mV}$. }
	\label{fig2}
\end{figure}

In order to obtain a better understanding of the surface termination and to understand why the topologically trivial surface states could be missing in the ARPES data, we performed STM experiments on the same batch of samples. Figure \ref{fig2} (a) shows a large scale topography image of a freshly cleaved sample. Several sharp step edges can be resolved which have two principal orientations rotated by 60$^\circ$ with respect to each other, indicating a good overall in-plane crystallinity. The corresponding histogram in Figure \ref{fig2} (c) shows that all step heights are integer multiples of $\approx0.75\,\mathrm{nm}$. This fits to the Tl-Tl distance (Figure \ref{fig2} (d)), also indicating that along the $z$-direction the sample shows the expected crystallinity. In contrast, the zoomed image in Figure \ref{fig2} (b) and the inset display a structure which resembles a partly ordered amorphous or liquid-like structure. We refer to the regions with higher apparent height in the inset of Figure \ref{fig2} (b) as worms. The same structure was observed over the complete sample surface regardless of sample and cleave, independent of cleaving temperature, scan parameters, and tip condition. These worms thus can be regarded as an intrinsic property of the cleaved TlBiSe$_2$ surface. In our STM measurements at different bias voltages we see no evidence of any dispersive electronic states within the worms, but it should be noted that based on these measurements alone we can not exclude the presence of such states.

\begin{figure}[tb]
	\centering
		\includegraphics[width=0.98\textwidth]{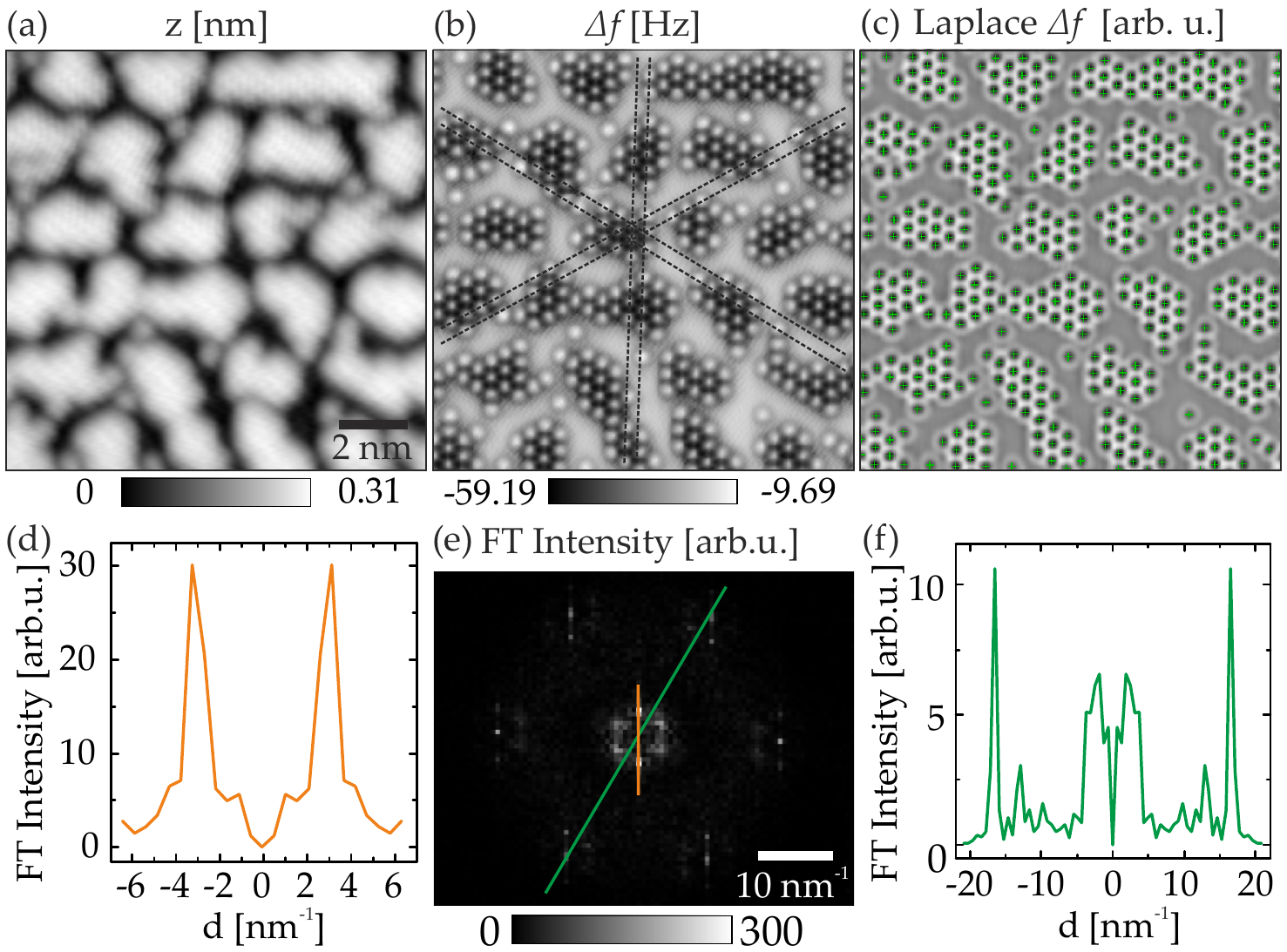}
	\caption{(a) Constant current STM topography data $(11.9\times11.9\,\mathrm{nm^2}$). (b) Constant height $\Delta f$ image of the same area as in (a). In the frequency shift image the hexagonal atomic structure within the worms is clearly resolved. (c) Low-pass and Laplace filtered version from (b), where each atomic site is marked with a cross \cite{Horcas:2007}. (d) Line profile from (e); the peaks are related to the periodicity of the worms. (e) Fourier spectrum of (b); the outer hexagon resembles the atomic ordering within the worms, the inner one the long-range hexagonal order of the worms. (f) Line profile from (e); the two peaks at $\pm 16.4\,\mathrm{nm^{-1}}$ are related to the periodicity of the atomic lattice. Imaging parameters: (a) $I=130\,\mathrm{pA}$, $U=200\,\mathrm{mV}$; (b) tip height $\Delta z = -230\,\mathrm{pm}$ with respect to the STM set point in (a), $U=10\,\mathrm{mV}$, $A=50\,\mathrm{pm}$, quality factor $Q=28140$, stiffness $k=1800\,\mathrm{N/m}$ and resonance frequency $f_0=26.666\,\mathrm{kHz}$.}
	\label{fig3}
\end{figure}

From our STM measurements it is not possible to determine the spatial extent of the worms normal to the surface, i.e. whether it is only one or several atomic layers. On the other hand, AFM measurements are able to resolve an atomic structure even if it is disordered \cite{Lichtenstein:2011}. Thus, we performed simultaneous STM/AFM measurements. The STM image obtained from this experiment (Figure \ref{fig3} (a)) closely resembles those measured with a dedicated STM setup shown in Figure \ref{fig2} (b), further supporting the universality of these results. It was not possible to obtain atomically resolved images in STM or AFM feedback mode. Therefore we switched to constant height mode, while gradually decreasing the tip sample distance, until atomic resolution within the worm-like structure showed up in the frequency shift $\Delta f$. In Figure \ref{fig3} (b) the tip was approached by $230\,\mathrm{pm}$ relative to the STM setpoint in Figure \ref{fig3} (a). Within the worms, a surprisingly large amount of local crystalline order was observed. A closer inspection reveals that, although no continuous connection is visible, all islands show the same crystal structure and orientation. Furthermore there is no shift in the registry of the atoms in different islands as indicated by the dashed lines in Figure \ref{fig3} (b). This is better visualized in the Fourier transform of the frequency shift data as shown in Figure \ref{fig3} (e), which displays two clear hexagonal patterns. The outer one is due to the atomic structure and the inner one is due to the hexagonal superstructure of the worms. Each of the outer peaks has satellites arising from the hexagonal superstructure of the worms. From the distance of the Fourier peaks, corresponding to the superstructure (Figure \ref{fig3} (d)) and the atomic structure (Figure \ref{fig3} (f)), the real-space average distance between the worms and the atomic lattice spacing are determined. For the superstructure we obtain $d_{\mathrm{worms}}=(2\pi)/(3.18\,\mathrm{nm^{-1}} \cdot \cos30^\circ)=2.28\,\mathrm{nm}$ and for the nearest-neighbor distance $d_{\mathrm{nn}}=(2\pi)/(16.30\,\mathrm{nm^{-1}} \cdot \cos30^\circ)=445\,\mathrm{pm}$, which is off by about $5\%$ from the bulk lattice constant of $425\,\mathrm{pm}$ \cite{Toubektsis:1987}. Note, that the direct hexagonal lattice is rotated by $30^\circ$ with respect to the reciprocal lattice. \newline
Although also here the second atomic layer can not be resolved, we conclude due to the well defined crystal structure within the worms and the correlation between them that only the topmost atomic layer is damaged during the cleaving process. This is further corroborated by the $30^\circ$ rotation of the superstructure peaks with respect to the atomically resolved structure in the Fourier spectrum which is expected for subsequent layers (Figure \ref{fig2}(d)). This suggests that the worms sit on a well ordered layer and all deeper layers have the expected crystal structure.

A quantitative analyis of Figure \ref{fig3} (b), performed by counting individual atoms, yields a number of 420 (Figure \ref{fig3} (c)). The total number of primitive units cells of area $A=\sqrt{3} d_{\mathrm{nn}}^2/2 $ which fit within the $11.9\times 11.9\,\mathrm{nm^2}$ scan area in Figure \ref{fig3} (b) is 826, resulting in a ratio of $420/826=0.51$. Apart from the atoms integrated in the islands we also observe several individual atoms in between. Such non-integrated atoms hint towards a composition of the worms of a metallic element and exclude a chalcogen such as Se. Furthermore, the lowest energy cleaving plane is found between the Tl and Se layers, where the distance between Tl and Se layers is $d_{\mathrm{TlSe}}=209\,\mathrm{pm}$ and $d_{\mathrm{BiSe}}=167\,\mathrm{pm}$ between Bi and Se layers \cite{Eremeev:2011Tl}, which suggests that the cleaving indeed occurs between Tl and Se layers. This is further corroborated by a recent XPS study which found a chemically different environment for the Tl atoms close to the surface and for those in the bulk of the crystal \cite{Kuroda:2013}, as also shown in Figure \ref{fig1} (c). 
Altogether we conclude that the surface is formed by Tl atoms with one half of the atoms remaining on the surface, while the other half is cleaved away. Due to the large amount of energy induced by the cleaving and the relatively weak bond between Tl and Se, the top atomic layer melts and then recrystallizes in the observed hexagonal superstructure even when the samples are cooled during the cleaving process. 

\begin{figure}[tb]
	\centering
		\includegraphics[width=0.98\textwidth]{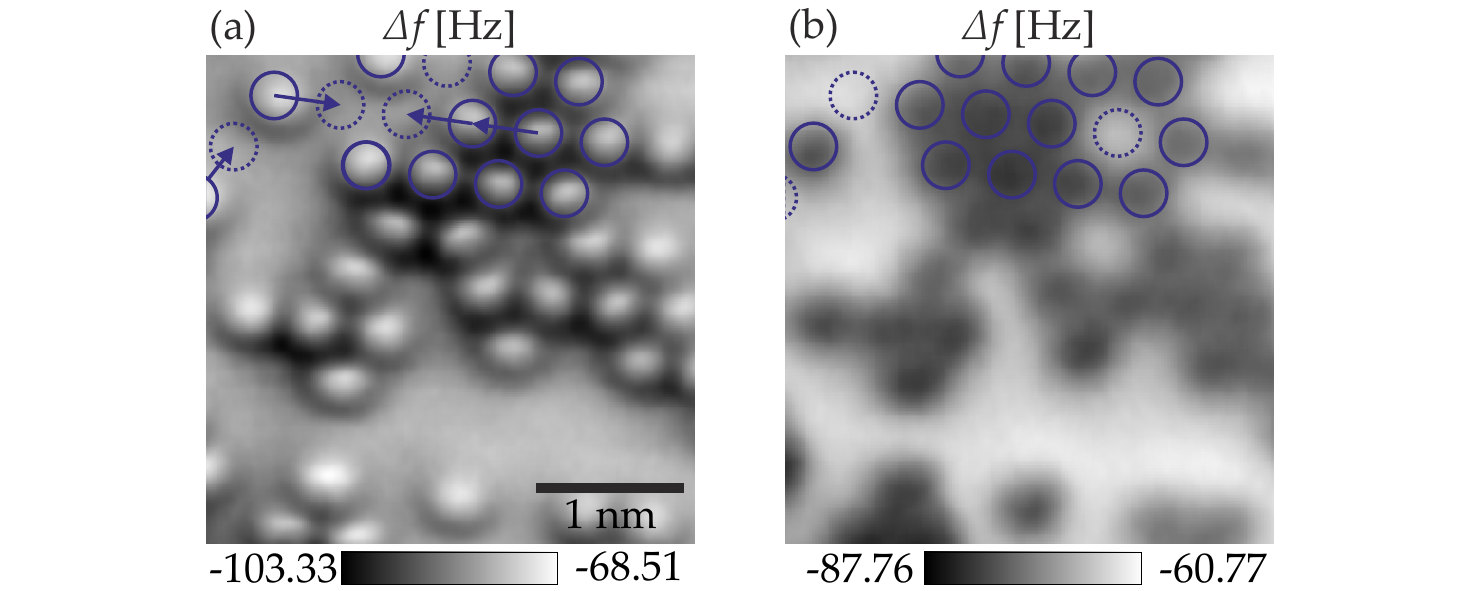}
	\caption{Atomically resolved images before (a) and after (b) a lateral manipulation process. The full (dashed) circles indicate occupied (unoccupied) atomic sites. The arrows mark the most likely manipulation path for the atoms. The different appearance of the atoms (bright in (a) and dark in (b)) is due to different relative tip-sample distances. In (a)  the tip is about $100\,\mathrm{pm}$ closer to the surface than in (b). Imaging parameters:  $U=10\,\mathrm{mV}$, $A=50\,\mathrm{pm}$, $Q=1.1\cdot 10^{6}$, $k=1800\,\mathrm{N/m}$ and $f_0=59.358\,\mathrm{kHz}$.}
	\label{fig4}
\end{figure}

 Figure \ref{fig4} shows two constant height $\Delta f$ images of the same area. The different appearance of the atoms is due to different relative tip-sample distances. Most notably, the positions of some of the atoms, which are indicated by blue circles, have changed in between the images. This is due to a lateral manipulation process induced by the tip. After Figure \ref{fig4} (a) was acquired the tip was approached closer to the surface while scanning in the upper region of Figure \ref{fig4} (a) until the oscillation amplitude became unstable. The tip was then retracted in constant height and the same area was imaged again (Figure \ref{fig4} (b)). A comparison of Figures \ref{fig4} (a) and (b) allows to identify that integrated (center atom of the hexagon) as well as non-integrated (top left atom) atoms were manipulated laterally. This suggest that the potential barrier for a lateral manipulation process is quite similar for integrated and non-integrated atoms and that the worms in itself are loosely bound.  

\begin{figure}[tb]
	\centering
		\includegraphics[width=0.98\textwidth]{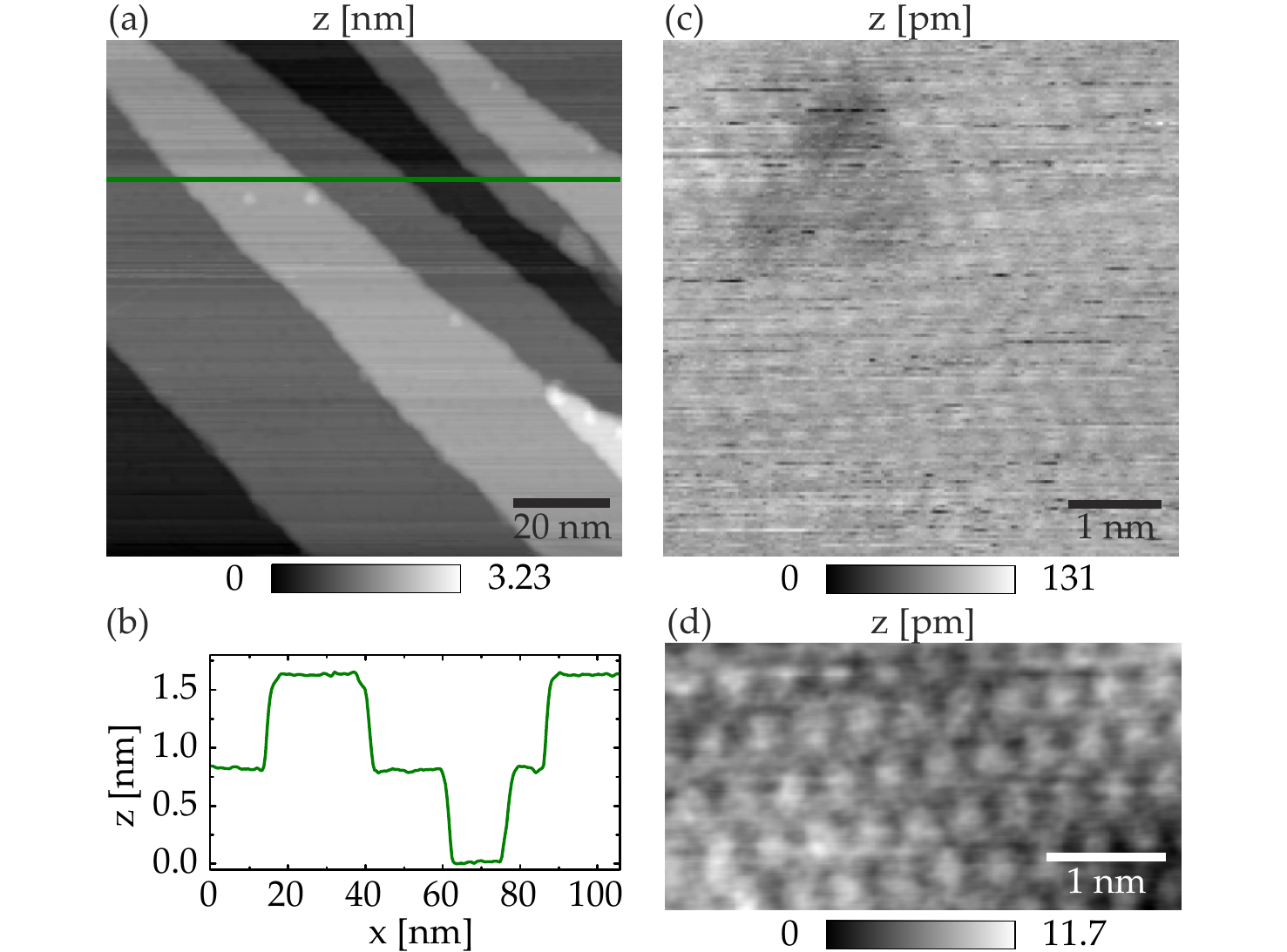}
	\caption{(a) Constant current STM topography data resolving several steps with heights of $\approx 0.8\,\mathrm{nm}$, see (b). (c),(d) The atomic resolution images show a regular hexagonal lattice. In (c) a triangular shaped defect can be identified. The image in (d) is low-pass filtered. The distortion of the hexagonal lattice is due to lateral drift and creep. Imaging parameters: (a) $I=100\,\mathrm{pA}$, $U=-800\,\mathrm{mV}$. (c) $I=50\,\mathrm{pA}$, $U=300\,\mathrm{mV}$. (d) $I=100\,\mathrm{pA}$, $U=-600\,\mathrm{mV}$.}
	\label{fig5}
\end{figure}

To answer the question how stable these islands are, we investigated the surface with STM at RT. Figure \ref{fig5} (a) shows steps with a height of about $0.8\,\mathrm{nm}$ (Figure \ref{fig5} (b)) which fits to the Tl-Tl distance. In contrast to our low temperature measurements we do not observe a worm-like structure. Instead, a regular hexagonal lattice is revealed in the atomically resolved images in Figures \ref{fig5} (c) and (d). Apart from the hexagonal lattice one can also identify a triangular depression in Figure \ref{fig5} (c) which we attribute to a sub-surface defect site similar as reported for Bi$_2$Se$_3$ \cite{Hor:2009,Cheng:2010}. We attribute the absence of the worms at RT to an increased mobility of the Tl atoms which move now too fast to be imaged by our slow STM (bandwidth $B\approx 1\,\mathrm{kHz}$). The large number of horizontal streaks in Figures \ref{fig5} (a) and (c) and the unstable imaging conditions support this further. Additionally, this interpretation is corroborated by studies of Tl on Si(111)--$7 \times 7$ at RT where Tl atoms are mobile on the surface and get trapped in an attractive potential to form nanodots \cite{Vitali:1999,Vitali:2001}. 
\section{Conclusion}
In the following we discuss the possible scenarios that combine our ARPES, STM, and AFM observations with published calculations \cite{Eremeev:2011Tl}. As mentioned above, calculations for all possible surface terminations show occupied topologically trivial spin-split surface states. Again, none of these states are observed in ARPES. The only exception is in case of a stacking fault at the surface resting in a -Se-Bi-Tl-Se structure instead of the expected -Se-Bi-Se-Tl unit. If such a stacking fault would be present it would result in a Se top layer, which is in direct conflict with our XPS and AFM results which indicate a Tl termination of the worms. Furthermore this interpretation would lead to the conclusion that in every sample studied by a variety of groups, grown in different laboratories, and for every cleave, a stacking fault is present exactly at the surface. Therefore, although we cannot exclude this possibility, it appears highly unlikely.

In Ref. \cite{Kuroda:2013} the authors gave a number of possible explanations for the absence of the trivial surface states in ARPES measurements. If dangling bond states exist but are localized on the small islands and in between them the contribution to photo-emitted electrons might be too small to be detected by ARPES. Other possible reasons are the ionic nature of the interlayer bonding between Tl and Se or the saturation of the dangling bonds due to a deformation of the islands. This deformation showed up as a reduced island height in the STM data of Ref. \cite{Kuroda:2013} compared to the bulk interlayer spacing. Our STM and AFM measurements at room and low temperature suggest that the crystalline islands form when the mobile Tl atoms freeze out during the cooling procedure of the sample. The two extreme situations would be either a huge Tl island which covers half of the cleavage surface or a uniform distribution of Tl atoms occupying each second lattice site. Intuitively one might think that in the latter case the dangling bonds of the underlying Se layer are most effectively saturated by the Tl atoms on top. The nanoscale islands (Figure 3(b)) which are formed consist of an average number of $16\pm6$ atoms. Furthermore only about $13\%$ of the atoms have six nearest neighbors and we found no atoms with six next-nearest neighbors within the surface plane. This suggests that the Tl atoms first prefer to stick together but once a certain island size is reached it is more favorable to form a new island. This would be in line with the above mentioned interpretation that the particular surface termination reduces the number of dangling bonds.

In an alternative, and more basic scenario, the crystallites on the surface are too small to allow for a Bloch-type wave to form and will thus not harbor any extended electronic states. Within this scenario, it is expected that any type of surface state is suppressed, which directly explains the absence of the spin-polarized termination-dependent topologically trivial surface states in the ARPES data. On the other hand, the topological surface state is clearly resolved in all ARPES measurements and appears not to be influenced by the surface structure. Lateral structures of similar dimensions result in the formation of quantum dots in the Cu(111) surface state \cite{Lobo-Checa:2009}. Due to the spin texture one would not expect the same simple quantisation mechanism for the TSS of TlBiSe$_2$ as not all scattering vectors are allowed. However, if the TSS were to have a significant probability density in the worms one would expect an influence on the measured spectra either in the form of broadening or quantisation effects. The absence of such effects in our or other published data, combined with the absence of the trivial surface states provides the first direct spectroscopic evidence for the topological protection of surface states on 3D topological insulators. This protection is not a consequence of the spin structure as the spin-polarized trivial surface states are destroyed, but follows directly from the transition from a topologically non-trivial to a trivial material \cite{Hasan:2010}. Because the top atomic layer can not form a well defined band structure this means by definition that it becomes topologically trivial. Therefore the topological transition occurs one layer or one stack deeper and the topological interface state is located there, while extending several unit cells into the bulk. This is very similar to how edge states move around defects in the quantum Hall effect. 

A similar protection mechanism was used to explain the observed surface state band structure of PbBi$_4$Te$_7$, but only indirect evidence could be provided there \cite{Eremeev:2012}. Furthermore, our results can help resolve the issue of whether mass acquisition and a small gap can occur at the Dirac point for sulfur doped (TlBiSe$_{2-x}$S$_x$) samples \cite{Sato:2011,Xu:2012arXiv}. Depending on the exact S concentration at the cleaving plane, the surface structure will differ and can in some case invoke extra scattering channels for small k-values. This is in line with the observation that whether a gap is found to occur varies from cleave to cleave \cite{Xu:2012arXiv}.

To conclude, through a combination of experimental techniques we have shown that only the topmost atomic layer of the topological insulator TlBiSe$_2$ is destroyed by the cleaving process. The crystalline Tl worms are too small to form a band structure and this layer therefore has a trivial topology. The interface between trivial and non-trivial band structure topology thus shifts towards the bulk and the topologically protected interface state forms here. This provides a direct explanation why the predicted trivial surface states are not observed with ARPES, but the spin-polarized topological interface state is. Consequently, the deliberate destruction of the surface can be a good method to suppress the occurrence of trivial surface states which could interfere with the desired topological transport properties.

\section{Acknowledgements}
Fruitful discussions with C. Mudry and T. Neupert are gratefully acknowledged. We thank A. Zeenny and V. Terrail for their help with the ARPES measurements. This work is supported by the Swiss National Science Foundation. F.P. and F.J.G. acknowledge financial support from the Deutsche Forschungsgemeinschaft (DFG) within SFB 689. J. W. acknowledges funding by the priority programme SPP1666 of the DFG and A.A.K. acknowledges funding from the Emmy Noether program of the DFG (KH324/1-1).


%

\end{document}